\documentclass[12pt]{article}
\usepackage{epsfig,exscale}
\usepackage{indentfirst}

\def\beq{\begin{equation}}
\def\eeq{\end{equation}}
\def\bea{\begin{eqnarray}}
\def\eea{\end{eqnarray}}
\def\bq{\begin{quote}}
\def\eq{\end{quote}}

\def\fr{\frac}
\def\ve{\varepsilon}
\def\dt{\Delta t}
\def\dm{\Delta m}
\def\dg{\Delta \Gamma}
\def\nn{\nonumber}

\newcommand{\hl}{\hline}
\newcommand{\ra}{\rightarrow}

\newcommand{\bo}{$B^0$}
\newcommand{\bb}{$\bar{B}^0$}
\newcommand{\ko}{$K^0$}
\newcommand{\kb}{$\bar{K}^0$}
\newcommand{\ks}{$K_S$}
\newcommand{\kl}{$K_L$}
\newcommand{\ol}{${\cal O}(\lambda^3)$}
\newcommand{\ie}{{\rm Im}(\varepsilon)}
\newcommand{\rd}{{\rm Re}(\delta)}
\newcommand{\rep}{{\rm Re}(\varepsilon)}
\newcommand{\id}{{\rm Im}(\delta)}
\newcommand{\jk}{$J/\Psi \  K_S$}
\newcommand{\jkl}{$J/\Psi \  K_L$}
\newcommand{\re}{{\rm Re}}
\newcommand{\im}{{\rm Im}}

\def\cp1sub{\setlength{\unitlength}{8pt}\begin{picture}(2,1)\mbox{\scriptsize CP} \end{picture}}

\def\gappeq{\mathrel{\rlap {\raise.5ex\hbox{$>$}}
{\lower.5ex\hbox{$\sim$}}}}

\def\lappeq{\mathrel{\rlap{\raise.5ex\hbox{$<$}}
{\lower.5ex\hbox{$\sim$}}}}

\title{Studying Indirect Violation of CP, T and CPT in a $B$-factory}

\begin{document}

\pagestyle{empty}
\begin{flushright}
FTUV-00-0530
\end{flushright}
\vspace*{15mm}
\begin{center}
{\bf \Large
Studying Indirect Violation of CP, T and CPT\\in a $B$-factory}\\

\vspace*{1.5cm}
{\bf  M. C. Ba\~ nuls} \\
\vspace*{0.3cm}
IFIC, Centro Mixto Univ. Valencia - CSIC, E-46100 Burjassot (Valencia), Spain.
\footnote{\tt banuls@ific.uv.es}\\
\vspace*{0.3cm}
and \\
\vspace*{0.3cm}
{\bf J. Bernab\'eu}\\
\vspace*{0.3cm}
Departamento de F\'{\i}sica Te\'orica, Univ. Valencia, E-46100
Burjassot (Valencia), Spain. 
\footnote{\tt Jose.Bernabeu@uv.es}\\
\vspace{0.3cm}
\vspace*{2cm}
{\bf ABSTRACT} \\ \end{center}
\vspace*{5mm}

In this work we analyze the observable asymmetries one can build 
from entangled $B$-meson states, in order to extract information on the 
parameters $\ve$ and $\delta$
which govern indirect violation of discrete symmetries.
The traditionally proposed observables, based on flavour tags,
are not helpful for the study of the $B_d$-system, where the tiny 
value of $\dg$ clears up such asymmetry effects.
Our study makes instead use of CP tags in order to build new asymmetries 
where the different parameters can be separated out.
For this separation, it is decisive to achieve a good time resolution
in the measurement of entangled state decays.
Nevertheless, even with no temporal information, as would be the case in
a symmetric factory, it is still possible to extract some information
on the symmetries of the system.
We discuss both non-genuine and genuine observables, depending on 
whether absorptive parts can mimic or not asymmetry effects.


\vfill\eject

\setcounter{page}{1}
\pagestyle{plain}

\section{Introduction}

The concept of indirect violation of discrete symmetries in a
neutral meson system makes reference
to the non-invariance properties of the effective hamiltonian
which governs the time evolution of mesons~\cite{ka68}.
Those properties can be analyzed by studying the symmetries 
in the problem of mixing during the time evolution of meson states, 
while the possible effects in direct decays must be excluded.
In these phenomena of time evolution, T violation is evidenced if,
for a given time interval, one detects a difference between the 
probabilities for $i \ra f$ and $f \ra i$, being $i$ and $f$
shortnames for neutral meson states.
CPT violation would be seen as a difference between $i \ra f$ 
and $\bar{f} \ra \bar{i}$.

For the kaon system, this study has been performed by the CP-LEAR 
experiment~\cite{cplear} from the preparation of 
definite flavour states \ko-\kb. 
These tagged mesons evolve in time and their later decay to
a semileptonic final state projects them again on a definite 
flavour state.
The study of this {\it flavour-to-flavour} evolution 
allows the construction of
observables which violate CP and T, or CP and CPT.
The results are interpreted in terms of non-invariances of 
the effective hamiltonian, although some discussion 
remains~\cite{ag98} on the possible CPT violation in 
semileptonic decays.

Nevertheless, these T- and CPT-odd observables would be zero, 
even in presence of T and CPT fundamental violation, 
if there were no absorptive components in the effective hamiltonian.
For the kaon system, the different lifetimes of physical states, \ks
~and \kl, ensures this is not the case.
On the contrary, in the case of the $B_d$-system, 
the width difference $\dg$ between the physical states
is expected~\cite{kh87} to be negligible.
Then the T- and CPT-odd observables proposed for kaons,
which are based on flavour tag, vanish for a $B$-meson system
with $\dg =0$.
But the $B_d$ entangled states can be used to
construct alternative observables which are sensitive to
T and CPT independently of the value of $\dg$~\cite{bb99.2}.

In this paper we study three different types of observables 
that can be constructed from the entangled states of $B_d$ mesons
in order to study indirect violation of CP, T and CPT.
We separate the observables into two different cathegories
\begin{enumerate}
\item
\emph{genuine} asymmetries, characterized by the fact that they are pure
symmetry observables, constructed by comparing the probabilities of two 
conjugated processes, so that the asymmetry vanishes if the relating
symmetry is conserved;
\item
\emph{non-genuine} asymmetries, which do not correspond to purely conjugated
pairs of processes, so that its non-vanishing value can be mimic by the
presence of absorptive parts.
\end{enumerate}
Within the first cathegory we can distinguish two types of observables
depending on whether they need the support of absorptive parts in order
to give non-vanishing asymmetries.

To start with we give in section~\ref{sec:intro} a brief overview 
of the formalism and notation used to study the problem of indirect 
violation of discrete symmetries in the $B$-system. 
Then, in section~\ref{sec:flav}, we will review how to construct
the flavour asymmetries, the analogous in the $B$-system to 
those observables measured for kaons. 
They turn out to be proportional to $\dg$, 
and their value is expected to be so small that they will 
not be useful to study the symmetry properties in the case
of the $B_d$-system.
They constitute the first type of observables we are studying, and belong
to the \emph{genuine} cathegory.

In section~\ref{sec:CP} we construct alternative asymmetries, 
based on a CP tag.
In that way we find CP-, T- and CPT-odd time-dependent asymmetries, 
whose values do not cancel even for $\dg=0$. 
This will be the second type of observables in our classification, 
offering the best chances to study the symmetries in $B_d$-mixing.
The resulting asymmetries are also \emph{genuine} observables.

The limit of small $\dg$ 
 causes the time-reversal operation and the exchange of decay products
to be equivalent.
We exploit this fact in section~\ref{sec:jks} to build the
third kind of asymmetries that involve only the \jk ~final state, 
experimentally easy to identify. 
These observables are \emph{non-genuine}, since their equivalence to the 
true T- and CPT-odd asymmetries only holds in the exact limit
$\dg=0$.

Therefore in section~\ref{sec:dg} we will see how the introduction 
of $\dg \neq 0$ affects the results in sections~\ref{sec:CP} 
and~\ref{sec:jks}, and we will discuss whether it is 
still possible to extract information on the CP, T and CPT parameters 
from the non-genuine asymmetries even in presence of fake effects due
to absorptive parts.

We emphasize that the time dependence of the observables 
plays a relevant role in order to separate genuine and fake effects.
But it is still possible to extract some limited information 
with measurements that have no time resolution, as we will discuss in 
section~\ref{sec:nodt}.

Finally, in section~\ref{sec:con} we will summarize our conclusions.

\section{The entangled state of $B$-mesons}
\label{sec:intro}

In a $B$ factory operating at the $\Upsilon(4S)$ peak, 
correlated pairs of neutral $B$-mesons are produced through the reaction:
\beq
e^+ e^- \ra \Upsilon(4S) \ra B \bar{B}. \nn
\eeq
In the CM frame, the resulting $B$-mesons travel in opposite directions, 
and each one will evolve with the effective hamiltonian of the 
neutral mesons system.
Due to the intrinsic spin of $\Upsilon(4S)$, the so formed state 
has definite $L=1$.
Bose statistics requires the physical \bo-\bb ~state 
to be symmetric under ${\rm C} \cal P$, being $\cal P$ the operator which 
permutes the spatial coordinates.
Together with this requirement, ${\rm C=-}$ implies that ${\cal P}=-$,
so that the initial state may be written as
\beq
\vert i>=\frac{1}{\sqrt{2}} 
\left (\vert B^0(\vec{k}), \overline{B}^0(-\vec{k})>
- \vert \overline{B}^0(\vec{k}), B^0(-\vec{k})> \right ).
\label{eq:ent}
\eeq

As a consequence, one can never simultaneously have two 
identical mesons at both sides of the detector.
This permits the performance of a flavour tag: if at $t=0$ 
one of the mesons decays through a channel which is only allowed 
for one flavour of the neutral $B$, the other meson in the pair 
must have the opposite flavour at $t=0$.
The correlation (\ref{eq:ent}) between both sides of the entangled 
state holds at any time after the production, until the moment 
of the first decay.

The entangled $B-\bar{B}$ state can also be expressed 
in terms of the CP eigenstates 
$|B_{\pm}\rangle\equiv\frac{1}{\sqrt{2}}\left(|B^0\rangle
\pm {\rm CP}|B^0\rangle \right )$ as
\beq
\vert i>=\frac{1}{\sqrt{2}} \left (\vert B_-(\vec{k}), B_+(-\vec{k})>
- \vert B_+(\vec{k}), B_-(-\vec{k})> \right ).
\label{eq:entCP}
\eeq

Thus, if the CP operator is well defined, it is also possible to 
carry out a CP tag.
It is enough for that to have a CP-conserving decay 
into a definite CP final state, 
so that its detection allows us to identify the decaying meson 
as a $B_+$ or a $B_-$.
As in the flavour case, it is possible to perform such a tag 
at any time after the production of the entangled state. 

In Ref.~\cite{bb99} we described how the determination of the CP operator 
is possible and unambiguous to \ol,
which is sufficient to discuss both CP-conserving and CP-violating
amplitudes in the effective hamiltonian for $B_d$ mesons.
Here $\lambda$ is the flavour-mixing parameter of the CKM matrix~\cite{ko73}.
The determination is based on the requirement of CP
conservation, to \ol, in the $(sd)$ and $(bs)$ sectors.
To this order, however, CP-violation exists in the $(bd)$ sector, 
and it can be classified by referring it to the CP-conserving direction.
A $B_d$ decay that is governed by the couplings of the
$(sd)$ or $(bs)$ unitarity triangles, or by the $V_{cd} V_{cb}^*$ side
of the $(bd)$ triangle, will not show any CP violation to \ol.
We may say that such a channel is free from direct CP violation.

Two complex parameters, $\ve_{1,2}$, 
describe the CP mixing in the physical states
\bea
|B_1 \rangle &=& \frac{1}{\sqrt{1+|\varepsilon_1|^2}} \left [|B_+
\rangle +
\varepsilon_1 |B_- \rangle \right ] \ , \nn \\
|B_2 \rangle &=& \frac{1}{\sqrt{1+|\varepsilon_2|^2}} \left [|B_-
\rangle +
\varepsilon_2 |B_+ \rangle \right ] \ .
\label{fis}
\eea
They are invariant under rephasing of the meson states, and physical 
when the CP operator is well defined~\cite{bb98}.

In case of CPT conservation, $\ve_1=\ve_2$.
There is another pair of parameters, $\ve$ and $\delta$, which can be 
alternatively used and has a simpler interpretation in terms of 
symmetries.
These parameters are defined as
\beq
\varepsilon \equiv \frac{\varepsilon_1 + \varepsilon_2}{2}, \hspace{2cm}
 \delta \equiv \varepsilon_1 - \varepsilon_2 \ .
\eeq
Their explicit expression in terms of the effective hamiltonian 
matrix elements,
when CPT violation is introduced perturbatively and
we may neglect terms which are quadratic in $\Delta \equiv H_{22}-H_{11}$,
 is given by
\beq
\ve=\fr{\sqrt{H_{12}CP_{12}^*}-\sqrt{H_{21}CP_{12}^{\phantom{*}}}}
{\sqrt{H_{12}CP_{12}^*}+\sqrt{H_{21}CP_{12}^{\phantom{*}}}} , \hspace{.5cm}
\delta=\fr{-2 \Delta}
{\left ( \sqrt{H_{12}CP_{12}^*}+\sqrt{H_{21}CP_{12}^{\phantom{*}}} \right )^2}.
\label{eq:param}
\eeq
The rephasing invariance
\footnote{We are using the notation $H_{i j}$, ${\rm CP}_{i j}$, etc. 
to represent the matrix elements of the corresponding 
operators in the flavour basis, for instance 
$H_{12} \equiv \langle B^0|H| \bar{B}^0 \rangle$.} 
of $\ve$ and $\delta$ is
apparent from Eq. (\ref{eq:param}).
In the following we will only keep linear terms in $\delta$.

When we pay attention to the restrictions imposed by discrete symmetries 
on the effective mass matrix, $H=M-\frac{i}{2}\Gamma$, we see that:
\begin{itemize}
\item{CP conservation imposes 
${\rm Im}(M_{12} {\rm CP}_{12}^*)={\rm Im}(\Gamma_{12}{\rm CP}_{12}^*)=0$ and
$H_{11}=H_{22}$;
\item{CPT invariance requires $H_{11}=H_{22}$}};
\item{T invariance imposes 
${\rm Im}(M_{12} {\rm CP}_{12}^*)={\rm Im}(\Gamma_{12}{\rm CP}_{12}^*)=0$}.
\end{itemize}
As a consequence, CPT invariance leads to $\Delta=0$ and thus $\delta=0$,
irrespective of the value of $\varepsilon$. 
Similarly, T invariance leads to $\varepsilon=0$,
independently of the value of $\delta$. 
CP conservation requires both $\varepsilon=\delta=0$.
Therefore we have four real parameters which carry information on 
the symmetries of the effective mass matrix, according to the following list
\begin{itemize}
\item{$\rep \neq 0$ signals CP and T violation, with $\dg \neq 0$;}
\item{$\ie \neq 0$ indicates CP and T violation;}
\item{$\rd \neq 0$ means that CP and CPT violation exist;}
\item{$\id \neq 0$ shows CP and CPT violation, with $\dg \neq 0$.}
\end{itemize}

To extract information on these symmetry parameters we may
study the time evolution of the entangled state (\ref{eq:ent}).
As can be found in the literature, the special features of 
this system can be used to extract
information on CP~\cite{wo84} and CPT~\cite{ko92} violation 
in $B$ mesons.

We use for the final state the notation $(X, \, Y)$, where $X$ 
is the decay product observed with momentum $\vec{k}$ at a time $t_0$, 
and $Y$ the product detected at a later time $t$ with momentum $-\vec{k}$.
The time variables $\dt=t-t_0$ (with definite positive sign) and $t'=t_0+t$ 
are used to describe the process instead of $t_0$ and $t$.
The probability to find an arbitrary final state $(X,\, Y)$ from the
initial state (\ref{eq:ent}) is given by
\bea
|( X,\, Y)|^2 &=& \frac{1}{8}
\left (\fr{1+|\ve_1|^2}{1-\ve_1 \ve_2}\right )^2
\left | \frac{1+ \varepsilon_2}{1+ \varepsilon_1} \right |^2
|\langle X | B_1\rangle |^2 |\langle Y | B_1\rangle |^2 e^{-{\Gamma} t'} 
\times \nonumber \\
& & \phantom {\times []} \times
\left \{
(\eta_+ + \eta_-)\cosh \left(\fr{\dg \dt}{2} \right ) - (\eta_+ - \eta_-) \cos (\Delta m \Delta t) \right. \nn \\
& & \phantom {\times [] +} 
\left.
+2 \eta_{\rm re} \sinh \left(\fr{\dg \dt}{2} \right )
-2 \eta_{\rm im} \sin (\Delta m \Delta t)
\right \} \ ,
\label{eq:prob}
\eea
where $\Gamma$ is the averaged width of the physical states $B_{1,2}$.
The $\eta$ coefficients are defined as
\bea
\eta_+ &=& |\eta_X + \eta_Y|^2, \nn \\
\eta_- &=& |\eta_X - \eta_Y|^2 , \nn \\
\eta_{\rm re} &=& {\rm Re} [(\eta_X +\eta_Y)(\eta_X^* - \eta_Y^*)] \ , \nn \\
\eta_{\rm im} &=& {\rm Im} [(\eta_X +\eta_Y)(\eta_X^* - \eta_Y^*)] \ ,
\label{eq:coef}
\eea
with 
\beq
\eta_X \equiv \frac{ \langle X |B^0 \rangle -
\frac{1-\varepsilon_2}{1+\varepsilon_2} {\rm CP}_{12}^*\langle X |\bar{B}^0
\rangle}
{\langle X |B^0 \rangle + \frac{1-\varepsilon_1}{1+\varepsilon_1} {\rm
CP}_{12}^*\langle
X |\bar{B}^0 \rangle}=
\frac{1+\varepsilon_1}{1+\varepsilon_2}\cdot
\frac{\varepsilon_2  \langle X |B_+\rangle+\langle X |B_-\rangle}
{\langle X |B_+\rangle+\varepsilon_1 \langle X |B_-\rangle}~,
\label{eq:etax}
\eeq
and an analogous expression for $\eta_Y$.
We write Eq. (\ref{eq:etax}) in terms of both flavour and CP eigenstates.
From the above expressions one can easily check that for $X=Y$
only $\eta_+$ remains and the probability $|( X,\, Y)|^2$ vanishes for
$\dt =0$.
This is a kind of EPR correlation, imposed here by Bose 
statistics~\cite{li68}. 
The integration of the probability (\ref{eq:prob}) over $t'$ between 
$\Delta t$ (always positive with our definition) and $\infty$ gives 
the intensity for the chosen final state, depending only on $\Delta t$
\beq
I(X, \, Y; \, \Delta t)=\fr{1}{2} 
\int_{\Delta t}^{\infty} d t' |(X, \, Y)|^2~.
\eeq

By comparing the intensities corresponding to different processes 
one builds time-dependent asymmetries that allow the extraction of
the relevant parameters.

\section{Flavour Tag: Genuine observables needing $\dg$}
\label{sec:flav}

As we mentioned in the introduction, when a semileptonic 
final state is seen at one side of the detector, the surviving meson
can be tagged as the conjugated flavour state.
Its later decay through another semileptonic channel is
equivalent to the projection of the evolved meson onto the corresponding
definite flavour state.
In our notation, we represent as  $\ell^{\pm}$ the final
decay product of a semiinclusive decay $B \ra \ell^{\pm} X^{\mp}$.
Thus, the final configuration we denote $(\ell, \, \ell)$ 
is equivalent to a {\it flavour $\ra$ flavour} evolution, at the
meson level.
In Table~\ref{tab:leptons} we show the equivalence for the four possible 
configurations with two charged leptons in the final state.
\begin{table}[h]
\begin{center}
\begin{tabular}{|c|c c c c|} 
\hl
Transition & $\bar{B}^0 \stackrel{\phantom 9}{\rightarrow}  B^0$ & 
$B^0 \ra \bar{B}^0$&
$\bar{B}^0 \ra \bar{B}^0$&
$B^0 \ra B^0$ 
\\ \hl
$(X,\, Y)$ & $(\ell^+, \, \ell^+ )$ &
$(\ell^-, \, \ell^-)$ &
$(\ell^+ , \, \ell^-)$ & 
$(\ell^- , \, \ell^+ )$ \\
\hl
\end{tabular}
\caption{Meson transitions corresponding to semileptonic $(X,\, Y)$ 
final decay products}
\label{tab:leptons}
\end{center}
\end{table}

From the processes of Table~\ref{tab:leptons} we may construct two 
non-trivial asymmetries. 
If we compare the intensities for the first and second processes 
and suppose there is no violation either of the $\Delta B=\Delta Q$ rule
or of CPT in the direct decay amplitudes,
we obtain the asymmetry
\beq
A (\ell^+, \, \ell^+) \equiv 
\fr{I(\ell^+,\ell^+)-I(\ell^-,\ell^-)}{I(\ell^+,\ell^+)+I(\ell^-,\ell^-)}=
\fr{1-\left| \fr{1-\ve_2}{1+\ve_2}\right|^2 
\left|\fr{1-\ve_1}{1+\ve_1}\right |^2}
{1+ \left| \fr{1-\ve_2}{1+\ve_2}\right |^2 
\left | \fr{1-\ve_1}{1+\ve_1}\right |^2}
\approx \fr{4 \fr{\rep}{1+ |\ve|^2}}{1 + 4 \left (\fr{\rep}{1+ |\ve|^2} \right )^2},
\label{eq:l+l+}
\eeq
where only linear terms in $\delta$ have been kept, since
CPT violation is treated in perturbation theory.
We observe that this Kabir asymmetry~\cite{ka68} is time independent.

This is a genuine CP and T asymmetry, since the second process 
corresponds to the CP-, or T-transformed of the first one.
Thus the asymmetry cannot be faked by absorptive parts in absence
of true T violation.
However, in the limit $\dg=0$ both $\rep$ and $\id$ vanish, and 
this quantity will be zero, even if CP and T violation exist.
So this observable, in spite of being CP- and T-odd, 
and constituting a true observable of CP and T violation, 
also needs, in order to be non-zero, the presence of $\dg \neq 0$.
This requirement is fulfilled in the kaon system, 
where the asymmetry (\ref{eq:l+l+}) has actually been measured
\cite{cplear}.
On the contrary, for $B_d$-mesons the negligible value of $\dg$
predicts that this asymmetry will be difficult to observe.
Present limits for $\rep$ are at the level of few percent~\cite{ac97}.

The second asymmetry to be constructed from semileptonic decays is~\cite{ko95}
\bea
A(\ell^+ , \, \ell^-) &\equiv& \frac{I(\ell^+,\ell^-)-I(\ell^-,\ell^+)}
{I(\ell^+,\ell^-)+I(\ell^-,\ell^+)}  
\nn \\
& \approx& - 2 \frac{{\rm Re} \left ( \frac{\delta}{1-\varepsilon^2}\right ) {\rm Sh} \frac{\Delta \Gamma \Delta t}{2} - {\rm Im} \left ( \frac{\delta}{1-\varepsilon^2}\right ) \sin(\Delta m \Delta t)}{{\rm Ch} \frac{\Delta \Gamma \Delta t}{2}+ \cos (\Delta m \Delta t)} ,
\label{eq:l+l-}
\eea
also to linear order in $\delta$.
This asymmetry, contrary to that in Eq. (\ref{eq:l+l+}), depends on time
as an odd function of $\dt$.
Eq. (\ref{eq:l+l-}) corresponds to a genuine CP and CPT asymmetry, as it compares the 
probabilities for $B^0 \ra B^0$ and $\bar{B}^0 \ra \bar{B}^0$.
These transitions are self-conjugated under time reversal transformation, 
so T violation is not operative in this case.
To get a non-zero value of the asymmetry (\ref{eq:l+l-}), 
both CP and CPT violation are required, but also $\dg \neq 0$. 
The proportionality to $\dg$ is present in both terms of the asymmetry, 
explicitly in the first one
and in the second term through $\id$.
Therefore measuring a small limit for
this observable does not give a straightforward bound on CPT violation, 
because the vanishingly small $\dg$ of $B$-mesons would 
hide any symmetry breaking effect.
Present limits~\cite{ac97} on $\id$ are again at the level of few percent.

\section{CP Tag: Genuine observables which do not need $\dg$}
\label{sec:CP}

As explicitly seen in Eqs.~(\ref{eq:l+l+}) and (\ref{eq:l+l-}),
the flavour asymmetries are difficult to prove
discrete symmetry violation in the $B$-system, 
due to the negligible $\dg$ between the physical states. 
However, we may construct alternative asymmetries 
making use of the CP eigenstates, 
which can be identified in this system by means of a CP tag.

From the entangled state (\ref{eq:entCP}) such a tag can be performed.
Let us suppose that $X$, a CP eigenstate produced along the CP-conserving 
direction, is observed at $t_0$ at one side of the detector.
Such a decay is free from direct CP violation,
and this assures that in the moment of the decay the surviving meson 
of the other side had the opposite CP eigenvalue.
One example of a decay channel with the properties
we are looking for is given by $X \equiv$ \jk, with CP$=-$, or by
$X \equiv J/\Psi \ K_L$, with CP$=+$, both of them governed by the 
``CP-allowed'' side $V_{cb}V_{cs}^*$ of the $(bs)$ triangle.
Then, the detection of such a final state leads to the preparation of the
remaining $B_d$ meson in the complementary CP eigenstate, to \ol.
Since CP is conserved in $K$ and $B_s$ systems to \ol, 
it is correct to identify the physical kaon states with the CP eigenstates, 
at least to this order in the expansion.

After this preparation, the CP tagged state is left to evolve for 
a certain time $\dt$.
We are interested in the transition probabilities from CP-eigenstates,
$B_+$ and $B_-$, 
to flavour states, \bo ~and \bb.
Thus we will be looking at final configurations such as 
$(J/\Psi \ K_S, \, \ell^+)$ that corresponds, 
at the meson level, to $B_+ \ra B^0$.
All the transitions connected to this one by symmetry transformations, 
are shown in Table~\ref{tab:jlep}.
\begin{table}[h]
\begin{center}
\begin{tabular}{|c|c c c|} \hl
Transition & 
$B_+ \stackrel{\phantom{-}}{\rightarrow} \bar{B}^0$ &
$\bar{B}^0 \stackrel{\phantom{-}}{\rightarrow} B_+$ &
$B^0 \stackrel{\phantom{-}}{\rightarrow} B_+$ 
\\ \hl
$\phantom{\stackrel{-}{-}}
(X,\, Y)
\phantom{\stackrel{-}{-}}$ & 
$($\jk, $\ell^-)$ & $(\ell^+$, \jkl $)$ &
$(\ell^-$, \jkl $)$ 
\\ \hl
Transformation & CP& CPT & T  \\ \hl
\end{tabular}
\caption{Transitions and $(X,\, Y)$ final configurations connected to $B_+
 \rightarrow B^0$ and $( J/\Psi \ K_S, \, \ell^+)$ by symmetry transformations.}
\label{tab:jlep}
\end{center}
\end{table}

Comparing the intensity of $(J/\Psi \ K_S, \, \ell^+)$ with each of the 
processes on Table~\ref{tab:jlep}, we construct three genuine asymmetries of the form
\beq
A(X,\, Y)=\fr{I(X,\,Y)-I(J/\Psi \  K_S,\, \ell^+)}
{I(X,\,Y)+I(J/\Psi \ K_S,\, \ell^+)}.
\eeq

Each of them is odd under the discrete transformation which
connects (\jk $, \, \ell^+)$ with $(X, \, Y)$.
To linear order in $\delta$ and in the limit $\dg =0$,
we get the following results:
\begin{itemize}
\item
The CP odd asymmetry,
\beq
A_{{\rm CP}}\equiv A(J/\Psi \ K_S, \, \ell^-)=
-2 \fr{\ie}{1+|\ve|^2} \sin (\dm \dt)
+\fr{1-|\ve|^2}{1+|\ve|^2} \fr{2 \rd}{1+|\ve|^2} 
\sin^2 \left (\fr{\dm\dt}{2}\right ),
\label{eq:aCP}
\eeq
has contributions from T-violating and CPT-violating terms.
The first term, odd in $\dt$, is governed by the T-violating $\ie$,
whereas the second term, which is even in $\dt$, is sensitive 
to CPT violation through the parameter $\rd$.
The CP asymmetry corresponds to the well known ``gold plate'' 
decay~\cite{bigi} and has been measured by CDF~\cite{CDF}.
The inclusion of CPT violation through $\rd$ has been 
considered previously in Ref.~\cite{ko95}.
The equivalence between final configurations and mesonic
transitions allows the interpretation of this observable
in terms of \emph{CP-to-flavour} transitions.
It constitutes then a test of indirect CP, either by T violation 
or by CPT violation.
The separation of T-odd and CPT-odd terms can be done by constructing
different asymmetries, which are discussed in the following.
\item
The T asymmetry,
\beq
A_{\rm T} \equiv A(\ell^-, \, J/\Psi \ K_L) =
-2 \fr{\ie}{1+|\ve|^2} \sin (\dm \dt)
\left[ 1 - \fr{1-|\ve|^2}{1+|\ve|^2} \fr{2 \rd}{1+|\ve|^2} 
\sin^2 \left (\fr{\dm \dt}{2} \right) \right],
\label{eq:aT}
\eeq
purely odd in $\dt$, needs $\ve \neq 0$, and includes CPT even 
and odd terms.
\item
The CPT asymmetry,
\beq
A_{{\rm CPT}} \equiv A(\ell^+, \, J/\Psi \ K_L)  =
\fr{1-|\ve|^2}{1+|\ve|^2} \fr{2 \rd}{1+|\ve|^2}
\fr{1}{1-2 \fr{\ie}{1+|\ve|^2} \sin (\dm \dt)}
\sin^2 \left (\fr{\dm \dt}{2}\right ),
\label{eq:aCPT}
\eeq
needs $\delta \neq 0$, and includes both even and odd time dependences,
so that there is no definite symmetry under a change of sign of $\dt$.
\end{itemize}

There is still a fourth discrete transformation, consisting
in the exchange in the order of appearance of the decay products
$X$ and $Y$, i. e. $\dt \ra -\dt$.
It transforms
\beq
\begin{array}{c}
B_+ \ra B^0 \\
(J/\Psi K_S, \, \ell^+)
\end{array}
\stackrel{\dt}{\ra}
\begin{array}{c}
\bar{B}^0 \ra B_- \\
(\ell^+, \, J/\Psi K_S)
\end{array}.
\label{eq:transdt}
\eeq
Looking at its effect at the meson level it turns out that $\dt$ reversal 
cannot be associated to any of the fundamental discrete symmetries.
In spite of this, it can still provide information on the symmetries 
of the system.
In the limit $\dg =0$, the temporal asymmetry satisfies
\beq
A_{\dt} \equiv A(\ell^+, \, J/\Psi \ K_S)=
  A(\ell^-, \, J/\Psi \ K_L) \equiv A_{\rm T} \ .
\label{eq:adt}
\eeq
This equality is a consequence of $\dg =0$. 
In general, the equivalence of T and $\dt$ inversions is only
valid for hamiltonians with the property of hermiticity, up to
a global (proportional to unity) absorptive part. 
For the kaon system, for instance, relation (\ref{eq:adt})
does not hold.

Measuring the presented asymmetries (\ref{eq:aCP})-(\ref{eq:adt})
with good time resolution, so to separate even and odd $\dt$ dependences,
should be enough to determine the parameters
\beq
\fr{2 \ie}{1+|\ve|^2}
, \hspace{2cm}
\fr{1-|\ve|^2}{1+|\ve|^2} \fr{2 \rd}{1+|\ve|^2},
\eeq
governing CP, T violation and CP, CPT violation, respectively, 
for the $B_d$ mixing.

Contrary to what happened in section~\ref{sec:flav}, the
CPT and T asymmetries based on a CP tag do not vanish due
to the smallness of $\dg$.
Instead, they provide a set of observables which could separate
the parameters $\delta$ and $\ve$.

In case of CPT invariance, $\delta=0$, $A_{\rm CPT}=0$, and 
the four asymmetries reduce to a single independent one 
$A_{\rm CP}=A_{\rm T}=A_{\dt}$.
In the Standard Model this measures $\sin (2 \beta)$ of the 
$(bd)$ unitarity triangle ~\cite{chk84}.
We emphasize, however, that one has access to four experimentally
different asymmetries whose results would also be different in case of 
CPT violation. 
In extended theories, efforts to find a plausible framework
for CPT violation are discussed in Ref.~\cite{ko99}.

\section{Non-genuine observables only with \jk}
\label{sec:jks}

All the asymmetries defined in the previous section are genuine
observables, since each of them compares the original process with
its conjugated under a certain symmetry and is thus odd under 
the corresponding transformation.
Nevertheless the measurement of all those quantities requires
to tag both $B_+$ and $B_-$ states.
The last needs, from the experimental point of view, a good reconstruction 
of the decay $B \ra $\jkl, not so easy to achieve as
for the corresponding \jk ~channel.
Therefore we may wonder how much can be learned about the symmetry 
parameters from the study of all possible asymmetries built
from final configurations $(X, \, Y)$ with only \jk.

We show that if one considers only final states in which a 
\ks ~is present, the same asymmetries can still be constructed,
provided that the limit $\dg =0$ is valid.
In this case there are four possible configurations of the 
final state, depending on the sign of the charged lepton 
and on the order of appearance of the decay products,
\beq
(J/\Psi K_S,\, \ell^+), \quad (J/\Psi K_S,\, \ell^-), \quad 
(\ell^+, \, J/\Psi K_S), \quad (\ell^-, \, J/\Psi K_S).
\label{eq:fin}
\eeq
We show in Table~\ref{tab:corr} the mesonic transitions which 
are related to each final state.
\begin{table}[h]
\begin{center}
\begin{tabular}{|c|c c c c|} 
\hl
Transition & $B_+ \stackrel{\phantom 9}{\rightarrow} B^0$ &
$B_+ \rightarrow \bar{B}^0$&
$\bar{B}^0 \rightarrow B_-$ &
$B^0 \rightarrow B_-$ \\ \hl
$(X,\, Y)$ & $(J/\Psi K_S, \, \ell^+ )$ &
$(J/\Psi K_S, \, \ell^-)$ &
$(\ell^+ , \, J/\Psi K_S)$ & 
$(\ell^- , \,J/\Psi K_S )$ \\
\hl
\end{tabular}
\caption{$(X,\, Y)$ configurations with \jk
~and the corresponding meson transitions}
\label{tab:corr}
\end{center}
\end{table}

In the exact limit $\dg=0$, taking into account that $\dt$ and T 
operations become equivalent, one can construct asymmetries which 
measure indirect violation of CP, T and CPT 
from the processes shown in Table ~\ref{tab:corr}.

Thus, we obtain, to linear order in the CPT violating parameter $\delta$,
\begin{itemize}
\item
\beq
A(J/\Psi K_S, \, \ell^-)
 =  2 \fr{\ie}{1+|\ve|^2} \sin (\dm \dt) 
+ 2 \fr{1-|\ve|^2}{1+|\ve|^2}\fr{\rd}{1+|\ve|^2} 
\sin^2 \left ( \fr{\dm \dt}{2} \right ).
\label{eq:asCP}
\eeq
This is the asymmetry $A_{\rm CP}$ defined in Eq. (\ref{eq:aCP}).
It measures $\rd$ or $\ie$ different from zero,
and constitutes indeed a genuine measurement of indirect CP violation.
Since it contains $\dt$ odd and $\dt$ even pieces, both 
parameters could be separated.
An equivalent observable would be given by the asymmetry between the third 
and fourth processes of Table ~\ref{tab:corr}, also related
by a CP transformation.
\item
\beq
A(\ell^+, J/\Psi K_S)
= - 2 \fr{\ie}{1+|\ve|^2} \sin (\dm \dt) \left [ 1-2 \fr{\rd}{1+|\ve|^2} 
\fr{1-|\ve|^2}{1+|\ve|^2}\sin^2 \left ( \fr{\dm \dt}{2}\right )\right ].
\label{eq:asT}
\eeq
This is the temporal asymmetry defined in the previous section.
It is only different from zero if $\ie$ is not null.
Then it measures indirect violation of T when $\dg=0$.
The second and fourth decays of the table are connected by a $\dt$ exchange, 
too, and will measure the same parameters.
\item
\beq
A(\ell^-, J/\Psi K_S)
= 2 \fr{\rd}{1+|\ve|^2} \fr{1-|\ve|^2}{1+|\ve|^2} 
\sin^2 \left ( \fr{\dm \dt}{2}\right ) \fr{1}{1- 2 \fr{\ie}{1+|\ve|^2} 
\sin (\dm \dt)}.
\label{eq:asCPT}
\eeq
This quantity is different from zero only if $\rd$ is not null.
It equals the CPT asymmetry $A_{\rm CPT}$ 
in Eq. ~\ref{eq:aCPT} and can be used to measure CPT violation if $\dg =0$.
\end{itemize}

These results point out a way of testing the discrete symmetries 
in the evolution of $B$-system, from data of entangled states decays. 
The observed events where a \ks ~is present in the final decay products 
can be classified according to Eq.(\ref{eq:fin}). 
This is enough to construct the different asymmetries sensitive 
to CP, T and CPT violating parameters $\ie$ and $\rd$, 
as long as $\dg=0$.
The method exploits the good $\dt$-time resolution accessible 
in asymmetric $B$-factories.
The separation of $\ve$ and $\delta$ is associated with resolving
odd and even functions of $\dt$, respectively.

\section{Linear $\dg$ corrections}
\label{sec:dg}

The asymmetries in 
Eqs. (\ref{eq:asT}) and (\ref{eq:asCPT}), contrary to those built in 
section~\ref{sec:CP}, are not genuine.
They do not correspond to true T- and CPT-odd observables, for
the processes we are comparing are not related by a symmetry
transformation. 
This implies that the presence of $\dg \neq 0$ may induce
non-vanishing values of the asymmetries discussed in section 
~\ref{sec:jks}, even if there is no true T- or CPT-violation.

Although in the limit $\dg=0$ the results of 
Eqs. (\ref{eq:asT}) and (\ref{eq:asCPT}) agree, respectively, with those 
in Eqs. (\ref{eq:aT}) and (\ref{eq:aCPT}), experimentally one is 
measuring different quantities.
When $\dg \neq 0$ is considered, 
absorptive parts affect all the above expressions, 
giving corrections to both genuine and non-genuine asymmetries.
These absorptive effects appear differently in observables
which yielded the same result before, and originate fake
contributions to non-genuine asymmetries.
To analyze them, we must take into account two types of corrections:
\begin{enumerate}
\item
time-dependent terms in the amplitudes which are governed 
by the evolution $\fr{\dg \dt}{2}$,
\item
non-vanishing values of $\rep$ and $\id$.
\end{enumerate}

To treat the problem in a systematic way, it is convenient to
study how the parameters $\rep$ and $\id$ depend on the non-vanishing
value of $\dg$.

In terms of the matrix elements of the effective hamiltonian, the 
relevant parameters were given by Eq. (\ref{eq:param}).
From these expressions we see
\beq
\rep \propto |M_{12}|^2 {\rm Im} \left (\frac{\Gamma_{12}}{M_{12}} \right ),
\quad
\id \propto \Delta |M_{12}| \left | \frac{\Gamma_{12}}{M_{12}}\right |.
\eeq

Since $\left | \frac{\Gamma_{12}}{M_{12}} \right |$ is proportional to $\dg$, 
and we are only interested in the first corrections to our asymmetries, we
may neglect $\id$, which is already first order in both $\dg$ and $\Delta$,
 and parametrize
\beq
\ve \equiv x \dg + i \ie, \quad \delta \equiv \rd.
\eeq

Using this parametrization it is now easy to throw away terms of 
order higher than one in $\dg$ or $\rd$.
The parameter $\ie$ will have also $\dg$ corrections. 
However, the new terms will 
accompany the zero order $\ie$
and have the same time dependence,
so that we will be unable to separate them.
Thus an explicit expansion of $\ie$ is not needed for our study.

\subsection{Corrections to flavour observables}

	We discussed in section~\ref{sec:flav} two asymmetries 
that could be built from flavour tag, Eq. (\ref{eq:l+l+}) and 
Eq. (\ref{eq:l+l-}).
Both of them vanished in case of $\dg=0$.
If we include linear $\dg$ corrections, as described above, the first
observable, $A(\ell^+, \, \ell^-)$, acquires a non zero value
\beq
A(\ell^+, \, \ell^+) \approx \fr{4 x \dg}{1+|\ve|^2},
\eeq
whereas the second one, $A(\ell^+, \, \ell^-)$, is linear in 
both $\dg$ and $\Delta$, and vanishes also at this order.

\subsection{Corrections to genuine \emph{CP-to-flavour} observables}

For the genuine asymmetries of section~\ref{sec:CP}, based
on a CP tag, we get
\begin{itemize}
\item
\bea
A_{\rm CP}&=&\frac{I(J/\Psi K_S, \ell^+)-I(J/\Psi K_S, \ell^-)}
{I(J/\Psi K_S, \ell^+)+I(J/\Psi K_S, \ell^-)} \nn \\
&\simeq& \frac{2 \rd}{1+|\ve|^2} \frac{1-|\ve|^2}{1+|\ve|^2}
\sin^2 \left (\frac{\dm \dt}{2} \right ) 
- \frac{2 \ie}{1+|\ve|^2} \sin (\dm \dt ) \nn \\
&&+\frac{4 x \dg}{1+|\ve|^2} 
\left [1+2 \left (\frac{2 \ie}{1+|\ve|^2} \right )^2 \right ]
\sin^2 \left (\frac{\dm \dt}{2} \right )
\nn \\ &&
-\frac{\dg \dt}{2} \frac{1-|\ve|^2}{1+|\ve|^2}
\frac{2 \ie}{1+|\ve|^2} \sin (\dm \dt) 
-\frac{8 \dg x}{1+|\ve|^2}\! \left (\frac{2 \ie}{1+|\ve|^2} \right )^2
\! \sin^4 \! \left ( \! \frac{\dm \dt}{2} \! \right ) \! .
\label{eq:dgCP}
\eea
With respect to Eq.(\ref{eq:aCP}), the $\dg$ corrections induce both
$\dt$ even and odd functions. 
The first two terms in Eq.(\ref{eq:dgCP}) reproduce the result for $\dg=0$,
whereas the third term has the same time dependence of the first one and
would make it difficult to extract $\rd$ from this CP asymmetry.
The last two terms have new $\dt$ dependences.
\item
\bea
A_{\rm T}&=&\frac{I(J/\Psi K_S, \ell^+)-I( \ell^-, J/\Psi K_L)}{I(J/\Psi K_S, \ell^+)+I(\ell^-, J/\Psi K_L)} \nn \\
&\simeq& -\frac{2 \ie}{1+|\ve|^2}\sin (\dm \dt)
\left [1- \frac{2 \rd}{1+|\ve|^2} \frac{1-|\ve|^2}{1+|\ve|^2} 
\sin^2 \left (\frac{\dm \dt}{2} \right )\right ] \nn \\
&&+\frac{4 x \dg}{1+|\ve|^2} \left[ 1-2 \left ( \frac{2 \ie}{1+|\ve|^2}\right )^2 \right ]\sin^2 \left (\frac{\dm \dt}{2} \right ) \nn \\
&&+\frac{\dg \dt}{2} \frac{1-|\ve|^2}{1+|\ve|^2}
\frac{2 \ie}{1+|\ve|^2} \sin (\dm \dt)+
\frac{8 x \dg}{1+|\ve|^2} \! \left (\frac{2 \ie}{1+|\ve|^2} \right )^2
\! \sin^4 \left (\! \frac{\dm \dt}{2} \! \right ) \!.
\label{eq:dgT}
\eea
Contrary to what happens for $A_{\rm CP}$, all the new terms in $A_{\rm T}$
from linear $\dg$ corrections have different $\dt$ dependences 
as compared to those in Eq.(\ref{eq:aT}).
\item
\bea
A_{\rm CPT}&=&\frac{I(J/\Psi K_S, \ell^+)-I( \ell^+, J/\Psi K_L)}{I(J/\Psi K_S, \ell^+)+I(\ell^+, J/\Psi K_L)} \nn \\
&\simeq&
-\frac{2 \rd}{1+|\ve|^2} \frac{1-|\ve|^2}{1+|\ve|^2}
\sin^2 \left (\frac{\dm \dt}{2} \right )
\fr{1}{1-\frac{2 \ie}{1+|\ve|^2}\sin (\dm \dt)}.
\label{eq:dgCPT}
\eea
To the order considered in our perturbation expansion, $A_{\rm CPT}$
has no linear $\dg$ corrections.
The genuine character of the asymmetry would put them in higher order
terms.
We conclude that, in presence of $\dg$ corrections, 
the extraction of $\rd$ should be done from $A_{\rm CPT}$.
\item
\bea
A_{\dt}&=&\frac{I(J/\Psi K_S, \ell^+)-I( \ell^+, J/\Psi K_S)}{I(J/\Psi K_S, \ell^+)+I(\ell^+, J/\Psi K_S)} \nn \\
&\simeq&
- 2 \frac{\ie}{1+|\ve|^2} \sin (\dm \dt) \left [ 1-2 \frac{\rd}{1+|\ve|^2} \frac{1-|\ve|^2}{1+|\ve|^2}\sin^2 \left ( \frac{\dm \dt}{2}\right )\right ] \nn \\
&& +\frac{\dg \dt}{2}\frac{1-|\ve|^2}{1+|\ve|^2} \nn \\ &&
-\frac{2 x \dg}{1+|\ve|^2}\frac{2 \ie}{1+|\ve|^2}\sin (\dm \dt)
\left [1-2\sin^2 \left (\frac{\dm \dt}{2} \right ) \right ].
\label{eq:dgdt}
\eea
From this equation we observe that the ``equivalence'' between 
time reversal and $\dt$ exchange does not hold any longer.
On the contrary, the difference between both asymmetries 
is now linear in $\dg$.
\end{itemize}

Except for $A_{CPT}$ in Eq. (\ref{eq:dgCPT}), all the asymmetries have 
linear corrections in $\dg$.
These terms make the analysis more complicated, 
but do not avoid the separation of parameters, provided one has 
enough statistics to identify the different time dependences 
in the asymmetries.
It is worth paying closer attention to the way linear 
$\dg$ corrections enter the different quantities.
$A_{\rm CP}$, $A_{\rm T}$ and $A_{\rm CPT}$ 
are genuine symmetry observables, i. e. they are
constructed from the comparison between processes 
related by a true symmetry transformation.
They are in general affected by corrections in $\dg$, 
but these do not generate fake effects such that 
even in absence of symmetry violation a non-vanishing asymmetry 
remains due to $\dg \neq 0$.
We can check that in expressions (\ref{eq:dgCP})-(\ref{eq:dgCPT})
all $\dg$ factors are multiplied either by $\ie$ or by $x$,
both vanishing in case of exact symmetries.
On the contrary, $A_{\dt}$, which as we have discussed does not 
correspond to any discrete symmetry, contains terms in $\dg$ 
when $\ve=\delta=0$.

\subsection{Corrections to non-genuine \emph{CP-to-flavour} observables}

The fact that $\dt$ and T inversions are no longer equivalent affects 
the non-genuine asymmetries we have built only with the ~\jk ~final state.
If we study how the linear terms modify the expressions in
Eqs.~(\ref{eq:asCP})-(\ref{eq:asCPT}), we find
\begin{itemize}
\item
the asymmetry in Eq. (\ref{eq:asCP}) is still a genuine CP asymmetry, 
and takes the same value as Eq.~(\ref{eq:dgCP});
\item
the asymmetry of Eq. (\ref{eq:asT}) is the temporal asymmetry, 
so it will be equal to that of Eq. (\ref{eq:dgdt}),
and cannot be identified with $A_{\rm T}$, as was the case for $\dg =0$;
\item
in the same way, Eq. (\ref{eq:asCPT}) is not equal to the true CPT 
asymmetry. Instead we get
\bea
A(\ell^- \!\!\!\!\! &,& \!\! J/\Psi K_S) \simeq 
\fr{1}{1-\frac{2 \ie}{1+|\ve|^2}\sin (\dm \dt)}
\left [
\frac{2 \rd}{1+|\ve|^2} \frac{1-|\ve|^2}{1+|\ve|^2} 
+\frac{\dg \dt}{2} \frac{1-|\ve|^2}{1+|\ve|^2}
\right. \nn \\
&&+\left.
\frac{4 x \dg}{1+|\ve|^2} \sin^2 \left ( \frac{\dm \dt}{2} \right )
-\frac{2 \ie}{1+|\ve|^2} \frac{2 x \dg}{1+|\ve|^2} \sin (\dm \dt)
\right ],
\eea
\end{itemize}
which contains linear terms in $\dg$ even when $\ve=\delta=0$.

\section{What can be seen in a symmetric factory?}
\label{sec:nodt}

In previous sections we saw the crucial role played by $\dt$ 
resolution in order to separate the symmetry parameters 
of the $B$-system.
To achieve the measurement of all the proposed observables
one needs to distinguish between configurations where final decay 
products $X$ and $Y$ occur with opposite time ordering,
and also to determine the time dependence of the resulting quantities
with good enough precision.

In symmetric $e^+ e^-$ facilities at the $\Upsilon(4S)$ peak, 
$B$ mesons are created at rest, so that they decay essentially
at the point of production. Therefore the measurement has no time 
resolution, but we may still extract some information 
on indirect symmetry violation from the available observations.

\subsection{Flavour Tag}

If $\dt$ is not measured, there are three different probabilities, 
\bea
\Gamma_{++}&=& \int_{0}^{\infty} d(\dt) \ I(\ell^+, \, \ell^+)=
\frac{C}{4 \Gamma} \left [ 1+\fr{4 x \dg}{1+|\ve|^2}+
\fr{2 \rd}{1+|\ve|^2}\right ] 
\fr{\left (\fr{\dm}{\Gamma} \right )^2}{1+\left(\fr{\dm}{\Gamma} \right)^2},
\nn \\
\Gamma_{--}&=& \int_{0}^{\infty} d(\dt) \ I(\ell^-, \, \ell^-)=
\frac{C}{4 \Gamma} \left [ 1-\fr{4 x \dg}{1+|\ve|^2}+
\fr{2 \rd}{1+|\ve|^2}\right ] 
\fr{\left (\fr{\dm}{\Gamma} \right )^2}{1+\left(\fr{\dm}{\Gamma} \right)^2},
\nn \\
\Gamma_{+-}&=& \int_{0}^{\infty} d(\dt) 
[I(\ell^+, \, \ell^-)+I(\ell^-, \, \ell^+)]=
\frac{C}{2 \Gamma} \left [ 1+\fr{2 \rd}{1+|\ve|^2}\right ] 
\fr{2+\left (\fr{\dm}{\Gamma} \right )^2}{1+\left(\fr{\dm}{\Gamma} \right)^2},
\label{eq:symll}
\eea
where $C$ is a common normalization factor for all the probabilities.

The only symmetry observable which can be constructed from these 
quantities is the asymmetry between $\Gamma_{++}$ and $\Gamma_{--}$. 
This yields the same result of Eq. (\ref{eq:l+l+}), since all
time dependence factored out from $I(\ell^+, \, \ell^+)$ and 
$I(\ell^-, \, \ell^-)$.

\subsection{CP Tag}

When we sum over all possible values of $\dt$, probabilities 
for channels which do not correspond to the same meson 
transition are added up.
For instance, $( J/\Psi \ K_S, \, \ell^+)$ and $(\ell^+, \  J/\Psi \ K_S)$,
which correspond to $B_+ \ra B^0$ and $\bar{B}^0 \ra B_-$,
contribute to the same probability.
In this way we may construct four quantities
\bea
\Gamma_{S+} & \equiv & \int_{0}^{\infty}
[I(J/\Psi K_S, \ell^+)+I(\ell^+, J/\Psi K_S)]d(\dt)  \nn \\
&&=\fr{C}{\Gamma} \left \{1+\fr{2 x \dg}{1+|\ve|^2} 
\fr{\left (\fr{\dm}{\Gamma} \right )^2}{1+\left(\fr{\dm}{\Gamma} \right)^2} 
+ \fr{\rd}{1+|\ve|^2}\left [ 2+ \fr{1-|\ve|^2}{1+|\ve|^2}
\fr{\left (\fr{\dm}{\Gamma} \right )^2}{1+\left(\fr{\dm}{\Gamma} \right)^2}\right ] \right \} ; \nn \\
\Gamma_{S-} & \equiv & \int_{0}^{\infty}
[I(J/\Psi K_S, \ell^-)+I(\ell^-, J/\Psi K_S)]d(\dt)  \nn \\
&&=\fr{C}{\Gamma} \left \{1-\fr{2 x \dg}{1+|\ve|^2} 
\fr{\left (\fr{\dm}{\Gamma} \right )^2}{1+\left(\fr{\dm}{\Gamma} \right)^2} 
+ \fr{\rd}{1+|\ve|^2}\left [ 2- \fr{1-|\ve|^2}{1+|\ve|^2}
\fr{\left (\fr{\dm}{\Gamma} \right )^2}{1+\left(\fr{\dm}{\Gamma} \right)^2}\right ] \right \} ; \nn \\
\Gamma_{L+} & \equiv & \int_{0}^{\infty}
[I(J/\Psi K_L, \ell^+)+I(\ell^+, J/\Psi K_L)]d(\dt) = \nn \\
&&=\fr{C}{\Gamma} \left \{1+\fr{2 x \dg}{1+|\ve|^2} 
\fr{\left (\fr{\dm}{\Gamma} \right )^2}{1+\left(\fr{\dm}{\Gamma} \right)^2} 
+ \fr{\rd}{1+|\ve|^2}\left [ 2- \fr{1-|\ve|^2}{1+|\ve|^2}
\fr{\left (\fr{\dm}{\Gamma} \right )^2}{1+\left(\fr{\dm}{\Gamma} \right)^2}\right ] \right \} ; \nn \\
\Gamma_{L-} & \equiv & \int_{0}^{\infty}
[I(J/\Psi K_L, \ell^-)+I(\ell^-, J/\Psi K_L)]d(\dt) = \nn \\
&&=\fr{C}{\Gamma} \left \{1-\fr{2 x \dg}{1+|\ve|^2} 
\fr{\left (\fr{\dm}{\Gamma} \right )^2}{1+\left(\fr{\dm}{\Gamma} \right)^2} 
+ \fr{\rd}{1+|\ve|^2}\left [ 2+ \fr{1-|\ve|^2}{1+|\ve|^2}
\fr{\left (\fr{\dm}{\Gamma} \right )^2}{1+\left(\fr{\dm}{\Gamma} \right)^2}
\right ] \right \} .
\eea

Each one of these probabilities corresponds to a pair of meson transitions, 
according to Table ~\ref{tab:sym}.
\begin{table}[h]
\begin{center}
\begin{tabular}{|c|c @{~or~} c | }
\hline
Probability & \multicolumn{2}{c|}{Transitions}\\
\hline
$\Gamma_{S+}$ & $B_+ \ra B^0$ & $\bar{B}^0 \ra B_-$ \\
$\Gamma_{S-}$ & $B_+ \ra \bar{B}^0$ & $B^0 \ra B_-$ \\
$\Gamma_{L+}$ & $B_- \ra B^0$ & $\bar{B}^0 \ra B_+$ \\
$\Gamma_{L-}$ & $B_- \ra \bar{B}^0$ & $B^0 \ra B_+$ \\
\hline
\end{tabular}
\end{center}
\caption{Mesonic processes associated to measurable probabilities 
without time resolution.}
\label{tab:sym}
\end{table}
Nevertheless we may construct genuine asymmetries by comparing the 
probabilities for conjugated pairs of transitions.
In this way we find
\bea
{\cal A}_{\rm CP} &\equiv& 
\fr{\Gamma_{S+}-\Gamma_{S-}}{\Gamma_{S+}+\Gamma_{S-}}
\approx \left [\fr{2 x \dg}{1+|\ve|^2}
+\fr{\rd}{1+|\ve|^2}\fr{1-|\ve|^2}{1+|\ve|^2} \right ] 
\fr{\left (\fr{\dm}{\Gamma} \right )^2}{1+\left(\fr{\dm}{\Gamma} \right)^2},
\label{eq:symCP}
\\
{\cal A}_{\rm T} &\equiv& 
\fr{\Gamma_{S+}-\Gamma_{L-}}{\Gamma_{S+}+\Gamma_{L-}}
\approx \fr{2 x \dg}{1+|\ve|^2}
\fr{\left (\fr{\dm}{\Gamma} \right )^2}{1+\left(\fr{\dm}{\Gamma} \right)^2},
\label{eq:symT}
\\
{\cal A}_{\rm CPT} &\equiv& 
\fr{\Gamma_{S+}-\Gamma_{L+}}{\Gamma_{S+}+\Gamma_{L+}}
\approx \fr{\rd}{1+|\ve|^2}\fr{1-|\ve|^2}{1+|\ve|^2} 
\fr{\left (\fr{\dm}{\Gamma} \right )^2}{1+\left(\fr{\dm}{\Gamma} \right)^2}.
\label{eq:symCPT}
\eea

Since no time dependence is measured, the distinction between
genuine and non-genuine observables, which was based on the 
equivalence between $\dt$ and T reversals for $\dg=0$, does not
apply in this case.
All  the asymmetries here are genuine, 
since none of them will give a non-zero value due to $\dg$ corrections,
unless there is also a violation of the corresponding symmetry.
Looking at the expressions, we realize that 
${\cal A}_{\rm CPT}$ is proportional to $\rd$, different
form zero only if CPT is not conserved.
${\cal A}_{\rm T}$ is proportional to the T-violating 
$\rep$, so that it needs also the presence of absorptive parts
to be detected.
Finally the CP asymmetry ${\cal A}_{\rm CP}$, already considered 
in the literature~\cite{ko95}, contains both a 
T-violating and a CPT-violating term, with no chance to separate
$\rep$ and $\rd$.
The most interesting result is therefore the access to a non-vanishing
value of $\rd$, even for $\dg=0$, from $A_{\rm CPT}$.

There is a constant factor 
$\fr{\left (\fr{\dm}{\Gamma} \right )^2}{1+\left(\fr{\dm}{\Gamma} \right)^2}$ common to all the asymmetries. 
\sloppy
For the $B_d$-system \mbox{$\fr{\dm}{\Gamma}=0.723 \pm 0.032$} \cite{lep99}, so that
this factor will not dilute appreciably the value of the observables.

\section{Conclusions}
\label{sec:con}

In this paper we have studied in detail the possibilities to explore 
indirect violation of discrete symmetries CP, T, CPT in a 
neutral meson system.
We have shown how, even in absence of relative absorptive parts, 
i. e., if $\dg=0$, this study is possible, but one needs
observables beyond {\it flavour-to-flavour} transitions due to
time evolution of the meson system.
In addition to flavour tags, then, we have considered also
CP tags, which are uniquely defined for the $B_d$-system at
order \ol, enough to include CP and T violation at 
leading order.
Possible CPT violation is included perturbatively.
This consistent scheme of the treatment of symmetry violation
can be tested at the $B$ factories, where the production
of entangled states of $B_d$ mesons allows the preparation
of the meson in a CP eigenstate.

The asymmetries analyzed in this work exploit their time 
dependences in order to separate out two different ingredients.
On one hand CP and T violation, described by $\ve$, and
on the other CP and CPT violation, given by $\delta$.
The complex parameters  $\ve$ and $\delta$ are defined in a 
rephasing invariant way and, due to the well defined CP 
operator, they are unique and physical quantities.

The observables we have described can be classified 
in three types:
\begin{enumerate}
\item
Genuine asymmetries for T or CPT violation, based on  
{\it flavour-to-flavour} transitions at the meson level,
which need the support of absorptive parts, $\dg \neq 0$.
The T asymmetry measures $\fr{\rep}{1+|\ve|^2}$ and
does not depend on time.
The CPT asymmetry is odd in $\dt$, but involves
$\re \left (\fr{\delta}{1-\ve^2}\right )$ and
$\im \left (\fr{\delta}{1-\ve^2}\right )$
with different time dependences.
Since both asymmetries are zero in absence of $\dg$, 
these are not promising observables for $B_d$ physics.

\item
Genuine observables which do not need $\dg$.
We have constructed three different asymmetries, 
signals of CP, CPT and T violation, respectively,
which are based on the combination of flavour and CP tags.
In the limit $\dg=0$ they involve $\fr{\ie}{1+|\ve|^2}$
and $\fr{1-|\ve|^2}{1+|\ve|^2}\fr{\rd}{1+|\ve|^2}$, 
which can be separated out from the different time
dependences.
There is yet a fourth quantity, the temporal asymmetry, experimentally 
different from the others, which becomes theoretically equivalent
to time reversal asymmetry only in the limit $\dg=0$.
In practice, this kind of observables needs the identification of 
a semileptonic decay in one side and \jk, \jkl ~decays in the other side.

\item
Making use of the equivalence between $\dt$ and T reversal operations
for $\dg=0$, we have also considered a group of non genuine observables.
They are equivalent to the genuine observables described in 
the previous paragraph in the limit $\dg=0$, with the advantage that 
they involve only the hadronic decay \jk. 
As a consequence, these asymmetries are accessible in the first
generation of $B$ factory experimental results.
These observables are non genuine in the sense that absorptive
parts can mimic a non-vanishing value for them, even if violation
of the fundamental symmetry is not present.

\end{enumerate}

The different character of genuine and non genuine observables
becomes apparent when corrections due to a non zero $\dg$ are included.
While in genuine observables the effect of $\dg$ is limited to correct
the existing terms of $\ve$ and $\delta$, non genuine asymmetries acquire
new terms, proportional to $\dg$, even for $\ve=\delta=0$.

In a symmetric $B$ factory, all the effects which are odd in $\dt$
are automatically cancelled by the integration over time 
implicit in all the experimentally measured quantities.
Nevertheless we have seen that some relevant information survives,
associated with even terms in $\dt$.

\section*{Acknowledgements}

\noindent

\mbox{M. C. B.} is indebted to the Spanish Ministry of Education and Culture for her fellowship.
This research was supported by CICYT, Spain, under Grant AEN-99/0692.

\end{document}